# Time- and Frequency-Resolved Observation of Inverse Orbital Hall Effect in Gallium Nitride via Terahertz Polarimetry

Kota Aikyo[1,*], Tomohiro Fujimoto[1,†], Ami Mi Shirai[1], Yuta Murotani[1,‡], Mitsuru Funato[2], Shinji Miwa[1], Jun Yoshinobu[1], Yoichi Kawakami[2], and Ryusuke Matsunaga[1,*]

[1]*The Institute for Solid State Physics, The University of Tokyo, Kashiwa, Chiba 277-8581, Japan*

[2]*Department of Electronic Science and Engineering, Kyoto University, Nishikyo-ku, Kyoto 615-8530, Japan*

**e-mail: aikyo-kota@issp.u-tokyo.ac.jp, matsunaga@issp.u-tokyo.ac.jp*

*†Present address: NTT Basic Research Laboratories, NTT Corporation, Atsugi, Kanagawa 243-0198, Japan.*

*‡Present address: Institute for Chemical Research, Kyoto University, Uji, Kyoto 611-0011, Japan*

## Abstract

**Orbitronics has attracted significant attention as a platform for information storage and processing that exploits the orbital angular momentum (OAM) of electrons without the need for spin-orbit coupling. However, experimental evaluation of OAM-charge interconversion remains a challenge and the results are often controversial due to the coexistence of bulk and interfacial contributions and the complexity in sample structures. Here, using circularly polarized light pulses and terahertz (THz) polarimetry, we developed a non-contact method to observe the inverse orbital Hall effect as a bulk response within a single material without employing heterostructure samples. In a semiconductor GaN, we directly captured the OAM-to-charge current conversion of holes in the THz frequency range. By analyzing the sharp frequency dependence of the Hall conductivity, we disentangled the competing microscopic mechanisms and revealed the dominant role of extrinsic contributions to the orbital Hall effect in the dc limit. By contrast, the Hall conductivity at THz frequencies above the impurity scattering rate is attributed to the intrinsic Berry-curvature mechanism,**

**allowing a quantitative comparison with the microscopic theory. Furthermore, the ultrafast dynamics of the inverse orbital Hall signal directly revealed sub-picosecond OAM relaxation of holes, comparable to that of phonon-mediated thermalization. The quantitative argument based on theoretical calculations suggested the existence of an even faster decay channel due to momentum redistribution by acoustic phonons, suggesting a sub-nanometer-scale OAM relaxation length. Our results provide comprehensive and crucial insights into OAM transport and establish an ultrafast, contact-free approach for investigating OAM-to-charge conversion.**

## I. INTRODUCTION

Electrical control of the internal degrees of freedom of charge carriers remains a central topic in fundamental and applied physics. Beyond the long-standing paradigm of spintronics [1], orbital angular momentum (OAM) of electrons has recently emerged as a novel controllable entity [2,3]. Of particular interest is the orbital Hall effect (OHE) [4-7], in which an orbital current—the flow of OAM—transverse to an applied bias field is generated in bulk crystals [Fig. 1(a)]. Although analogous to its spintronic counterpart, the spin Hall effect (SHE) [8], the OHE is distinct in that it emerges without relying on spin-orbit coupling (SOC) [7], thus enabling the construction of heavy-element-free devices.

A major challenge in studying the OHE (SHE) is that the orbital (spin) current cannot be directly observed in experiments. Fundamentally, the definition of orbital (spin) current itself remains under debate [9-15] because OAM (spin) is not a conserved quantity in solids. Experimental reports on the OHE have relied on indirect detection, such as OAM accumulation at sample edges [16-19] and OAM-induced torque in adjacent ferromagnets [16,20-25]. The involvement of surfaces and interfaces in such thin-film studies substantially complicates the attribution of the signals, owing to the interfacial Rashba-Edelstein effect [26-31]. Moreover, heterostructures containing ferromagnets suffer from the proximity effect that can modify the intrinsic magnetic properties of the materials [32]. The complexities inherent to heterostructures and the numerous assumptions underlying the analysis may be responsible for the discrepancies up to a few orders of magnitude in the reported orbital Hall conductivities [16,20,22,23] and OAM relaxation lengths [17,20,22,23,25,33].

In the context of its spin counterpart, these issues have been circumvented by an optical approach based on the inverse spin Hall effect (ISHE)—the conversion of a spin current into a transverse charge current; circularly polarized (CP) light generates spin-polarized photocarriers [34,35], and the population imbalance between the opposite spin states produces a transverse current under an electric field via the ISHE [36-39]. In such experiments, the input (spin polarization) can be quantitatively evaluated from optical selection rules, and the output (charge current) can be directly measured as a macroscopic anomalous Hall response. Since both represent well-defined bulk quantities, this approach enables a rigorous evaluation of spin-charge interconversion, allowing a comparison with microscopic theories without the complications present in standard SHE experiments. To capture this transport and its ultrafast dynamics, time-resolved spectroscopy employing CP pump and terahertz (THz) probe pulses has emerged as an ideal scheme [40-42]; the THz field drives spin-polarized carriers to generate an oscillating transverse current, the emission of which is measured as the Faraday rotation of the transmitted THz polarization. Owing to its contactless, transmissive nature and sub-picosecond time resolution, this technique is highly suitable for probing bulk responses and their temporal evolution, effectively eliminating artifacts from sample structures. Crucially, recent studies have also revealed that another nonlinear photocurrent, originating from asymmetric carrier excitation in momentum space under a bias field, substantially overlaps with the light-induced ISHE in the same experimental geometry [43-45]. To properly exclude this mechanism, termed the field-induced circular photogalvanic effect (FI-CPGE), ultrafast time-resolved detection is essential [45]. Accordingly, time-resolved THz polarimetry in GaAs has successfully observed its spin Hall conductivity spectrum and unambiguously quantified the intrinsic and extrinsic contributions [42].

This approach can be extended to investigate OAM transport because the conservation of angular momentum dictates that CP excitation primarily generates OAM polarization of carriers, inducing spin polarization as a secondary effect through SOC. Driving the OAM-polarized carriers with a bias field generates a measurable transverse charge current, constituting the inverse orbital Hall effect (IOHE) [Fig. 1(b)]. Previously, observations of the IOHE were reported using laser pulse-induced THz emission in nonmagnetic/ferromagnetic thin-film bilayers [33,46]; however, similar experiments [30,31] have attributed the THz emission to the inverse orbital Rashba-Edelstein effect rather than the IOHE, which underscores the significance of interfaces in such heterostructures. Notably, the results can be qualitatively distinct even for the very same film thicknesses in the heterostructure [30,33], implying their sensitivity to the surface quality. In stark contrast, time-resolved THz polarimetry utilizing CP pump pulses enables direct observation of the IOHE in a single bulk material and quantitative evaluation of the orbital Hall conductivity, thereby offering unprecedented insights into the OHE and the IOHE.

In this paper, we investigate the light-induced IOHE in a wide-gap semiconductor GaN. The weak SOC due to the light element N and the crystal field splitting of the wurtzite structure allow the direct gap at the $\Gamma$ point [Fig. 1(c)] to be well described in terms of OAM eigenstates. Figure 1(d) shows approximate optical selection rules at the zone center. Equally exciting all three valence bands with a CP pump induces OAM polarization of holes, while suppressing spin polarization of both holes and electrons. As shown in Fig. 1(e), our calculation based on the $\mathbf{k} \cdot \mathbf{p}$ method [47,48] also confirms that OAM ($L_z$) polarization is efficiently induced near the band edge, and it dominates over spin ($S_z$) polarization [49]. Furthermore, the weaker SOC [62] and the larger bandgap compared to typical spintronic semiconductors such as GaAs [63] imply that the intrinsic spin Hall conductivity is small. Therefore, GaN provides an excellent platform for the unambiguous observation of the light-induced IOHE. Using an ultraviolet (UV) CP pump and polarization-resolved THz spectroscopy in GaN, we present a comprehensive study of the IOHE in a time- and frequency-resolved manner, revealing the microscopic mechanisms underlying the frequency-dependent Hall conductivity as well as the ultrafast transport and relaxation dynamics of OAM.

## II. METHOD

### A. Time-resolved THz polarimetry

Our experimental setup is schematically illustrated in Fig. 2(a). The UV pump, THz probe, and near-infrared (NIR) gate pulses were generated from the output of a Yb-based regenerative amplifier with a 255-fs pulse duration and a 1028-nm center wavelength. The UV pump pulses were obtained by optical parametric amplification followed by two consecutive stages of second-harmonic generation, and their photon energy ($\hbar\omega_{\mathrm{pump}}$) was tuned from 3.31 to 3.57 eV around the band edge of GaN ($E_g =$ 3.39 eV). The pump pulse duration for above-bandgap excitation was evaluated to be 130–160 fs via autocorrelation, depending on the photon energy. The THz probe pulses were generated by optical rectification in a 2-mm-thick GaP$(110)$ crystal and were linearly polarized along the horizontal ($x$-) direction through a wire-grid polarizer (WGP0) before incidence on the sample. The transmitted THz field was detected by electro-optic sampling in another GaP$(110)$ crystal with the gate pulses compressed to 100 fs using the multiplate spectral broadening technique [64,65]. The horizontal and vertical components of the THz field, $E_x$ and $E_y$, were separately measured with two wire-grid polarizers, WGP1 and WGP2; $E_y$ was isolated by blocking $E_x$ with both polarizers, while $E_x$ was extracted from the average of the two polarization components, $(E_x + E_y)/2$, obtained with WGP1 rotated by 45° [66,67]. With the optical path of the gate pulses fixed, the UV pump delay $t_{\mathrm{pump}}$ and the THz probe delay $t_{\mathrm{probe}}$ were independently scanned by two mechanical stages. To properly extract the helicity-dependent polarization rotation due to the IOHE, the polarization of the UV pump was switched between left- and right-handed circular polarizations (LCP and RCP) by rotating a quarter-wave plate in the pump path. More details are described in Supplemental Material [49].

### B. Sample

The sample was a 10-μm-thick $(0001)$-oriented GaN film grown on a 430-μm-thick $(0001)$-oriented sapphire substrate by metal-organic chemical vapor deposition. Since the penetration depth of the pump (approximately 1 μm above the bandgap) was shorter than the film thickness, the signals were analyzed using a two-layer model [68]. This model approximates the sample as a homogeneous excited layer with a thickness equal to the penetration depth and an underlying unexcited layer. The UV pump and THz probe were normally incident on the sample, with the pump spot size (1.7 mm) set larger than that of the THz probe (0.5 mm at 1 THz) to ensure in-plane homogeneous excitation. All experiments were conducted at room temperature.

## III. RESULTS

### A. Time-domain analysis of THz field

Figure 2(b) shows the transmitted THz waveform $E_x(t_{\mathrm{probe}})$ in the absence of the pump pulse. A two-dimensional (2D) plot of the light-induced change in the THz field, $\Delta E_x(t_{\mathrm{pump}}, t_{\mathrm{probe}})$, measured at $\hbar\omega_{\mathrm{pump}} =$ 3.42 eV, is presented in Fig. 2(c). A profile along the $t_{\mathrm{probe}}$-axis yields a transient THz waveform indicating the longitudinal response, while the dynamics following photoexcitation are obtained from a profile along the $t_{\mathrm{pump}}$-axis. The decrease in the transmitted amplitude observed in Fig. 2(c) reflects the screening of the THz field by excited carriers. Fourier analysis of the transient THz waveforms before and after excitation yields the pump-induced change in the longitudinal complex conductivity spectrum, $\Delta\sigma_{xx}(\omega)$, over the THz frequency range [Fig. 2(d)]. The Drude model provides a good fit to the spectrum, with the carrier density $N_{\mathrm{car}}$ and longitudinal momentum relaxation time $\tau_{xx}$ estimated to be $1.2 \times 10^{17}$ cm$^{-3}$ and 0.15 ps, respectively. Furthermore, an analysis of the dynamics along the $t_{\mathrm{pump}}$-axis indicates that the carrier lifetime is on the order of 100 ps [49], well beyond the time window explored here.

Turning to the transverse response, Fig. 3(a) displays a 2D plot of the helicity-dependent THz field $\Delta E_y(t_{\mathrm{pump}}, t_{\mathrm{probe}}) \equiv \left(E_y^{\mathrm{LCP}}(t_{\mathrm{pump}}, t_{\mathrm{probe}}) - E_y^{\mathrm{RCP}}(t_{\mathrm{pump}}, t_{\mathrm{probe}})\right)/2$, which decays much faster than the long-lived longitudinal response in Fig. 2(c). Because this signal is largely confined to the vicinity of zero pump delay, a critical step here is to confirm that it genuinely originates from the anomalous Hall response of photoexcited carriers rather than the FI-CPGE; the latter is a third-order nonlinear current generation process that occurs only during the temporal overlap of the CP pump and the THz bias field. Importantly, our time-resolved 2D mapping unambiguously discriminates between the two mechanisms. Figure 3(b) shows the simulated 2D profile of the FI-CPGE (see Supplemental Material [49] for details). Since the time-domain response function of the FI-CPGE includes a delta function $\delta(t_{\mathrm{pump}} - t_{\mathrm{probe}})$, its contribution would manifest as a characteristic checkerboard pattern along the diagonal line $t_{\mathrm{pump}} - t_{\mathrm{probe}} = 0$ in the 2D plot [45]. However, the measured $\Delta E_y$ signal in Fig. 3(a) clearly extends along the $t_{\mathrm{pump}}$-axis. This feature demonstrates that the transverse current is indeed the anomalous Hall current of photocarriers; it can be observed even after the pump pulse has passed through the sample as long as the light-induced OAM (or spin) polarization persists, as simulated in Fig. 3(c) [49]. We found that the FI-CPGE in GaN is relatively weaker than that in a Dirac semimetal $Cd_3As_2$ [44] and GaAs [45]. This may be attributed to the valence-band splitting due to the crystal field and SOC, which prevents the inter-valence-band resonance for strong FI-CPGE [49]. Note that our result does not necessarily imply the negligibility of the FI-CPGE in the static Hall resistivity measurements using electrodes, since the detection efficiency of the FI-CPGE is highly suppressed in the time-resolved THz polarimetry compared to the contact-based transport experiments [45]; the primary component of the FI-CPGE signal is filtered due to the diffraction (see Supplemental Material for details [49]).

### B. Frequency-resolved orbital Hall conductivity

The transient THz waveform taken shortly after the arrival of the pump pulse, at $t_{\mathrm{pump}} = 0.3$ ps, allows us to extract the frequency dependence of the light-induced Hall conductivity, $\Delta\sigma_{yx}(\omega)$. Figures 3(d) and 3(e) show the real and imaginary parts of the Hall conductivity spectrum, respectively. The distinct dispersion in the spectrum is well reproduced by the solid curves showing our phenomenological model, which is constructed by analogy with the anomalous [69,70] and spin [42] Hall conductivities:

$$\Delta\sigma_{yx}(\omega) = \Delta\sigma_{yx}^{\mathrm{int}}\left(1+\frac{C_{\mathrm{sj}}}{1-i\omega\tau_{yx}}+\frac{\omega_{\mathrm{skew}}\tau_{yx}}{(1-i\omega\tau_{yx})^2}\right). \tag{1}$$

Here, the first term represents the intrinsic contribution related to the Berry curvature in momentum space, which is assumed to be constant over the probed frequency range below 10 meV. By contrast, the second and third terms are frequency-dependent extrinsic contributions from the side-jump and skew scattering mechanisms, the strengths of which relative to the intrinsic term are quantified by $C_{\mathrm{sj}}$ and $\omega_{\mathrm{skew}}$, respectively. $\tau_{yx} \sim$ 0.5 ps is the transverse impurity scattering time. The three contributions to the orbital Hall conductivity, constituting the fitting curves in Figs. 3(d) and 3(e), are separately plotted in Figs. 3(f) and 3(g). Notably, the overall behavior of the orbital Hall conductivity spectrum in GaN closely resembles its spin counterpart in GaAs [42]. Our observation of the sharp frequency dependence of the orbital Hall conductivity strongly indicates the dominant role of extrinsic mechanisms in the dc limit. Although earlier studies of the OHE have predicted its robustness against disorder [4] and most studies have focused on the intrinsic origin thus far, recent theories have suggested the importance of impurity scattering [71-73], which is in good agreement with our result. On the other hand, Figs. 3(f) also shows the prominence of the intrinsic response in the THz region, as the effects of impurity scattering are suppressed in the high-frequency regime ( $\omega/2\pi > \tau_{yx}^{-1}$ ). This demonstrates that the observation of the orbital Hall conductivity spectrum in the THz frequency range enables a direct comparison with the intrinsic contribution calculated from the band structure.

### C. Pump-photon-energy dependence of parameters

The $\hbar\omega_{\mathrm{pump}}$ dependence of the intrinsic Hall conductivity $\Delta\sigma_{yx}^{\mathrm{int}}$ extracted from the experiment is presented in Fig. 4(a). Since the spin lifetime of the *s*-like conduction electrons—tens of picoseconds at room temperature [74,75]—is much longer than that of the observed transverse response, we attributed the signal to the photoexcited holes. To compare this result with a microscopic theory, we calculated the light-induced anomalous Hall conductivity of holes based on the Kubo formula, given by

$$\Delta\sigma_{yx}^{\text{int}} = \frac{2e^3}{\hbar^3}\sum_{\mathbf{k}}\sum_{\substack{n\in\text{VB}\\ m\in\text{CB}\\ l\neq n}} \text{Im}\left[\text{tr}_n\left((\mathbf{A}^*_{\omega_{mn}}\cdot\mathbf{v})P_m(\mathbf{A}_{\omega_{mn}}\cdot\mathbf{v})\frac{P_n j_y P_l v_x}{\omega_{ln}^2}\right)\right]. \quad (2)$$

Here, $n$, $m$, and $l$ are the band indices for the Kramers-degenerate subspaces; specifically, VB and CB refer to the valence and conduction bands, respectively. We define $\omega_{mn} \equiv \omega_m - \omega_n$, where $\hbar\omega_n$ is the eigenenergy of band $n$. As gauge-invariant operators, $\text{tr}_n$ represents the trace over band $n$, while $P_n$ is the band projection operator. The quantity $\mathbf{A}_\omega$ is the vector potential of the CP pump pulse with frequency $\omega$. The velocity (**v**) and charge current (**j**) operators are given by $\hbar\mathbf{v} = \nabla_{\mathbf{k}} H_{\mathbf{k}}$ and $\mathbf{j} = -e\mathbf{v}$, with $H_{\mathbf{k}}$ being the $\mathbf{k}\cdot\mathbf{p}$ effective Hamiltonian defined in a **k**-independent basis.

The observed Hall current fundamentally comprises both the IOHE and ISHE contributions. Although a naive decoupling of the orbital and spin components breaks down—since even weak SOC prevents OAM and spin from being strictly good quantum numbers—we have developed an analytically exact formalism. In this framework, the measured Hall conductivity is written in terms of the orbital and spin Hall conductivities normalized by the carrier polarization, for which we rely on the total angular momentum, $J_z = L_z + S_z$, as a well-defined quantity for the electronic states of GaN. At the $\Gamma$ point, we have $L_z^\Gamma = \hbar(|x+iy\rangle\langle x+iy| - |x-iy\rangle\langle x-iy|)$ with $|x\rangle$ and $|y\rangle$ denoting cell-periodic basis states for the $\Gamma_5$ bands. For finite **k,** we need an additional term to account for the intercell contribution; setting $L_z \equiv L_z^\Gamma + \hbar(xk_y - yk_x)$ preserves the commutativity of $J_z$ and $H_{\mathbf{k}}$. Using the definitions above, we obtain

$$\Delta\sigma_{yx}^{\text{int}} = -\frac{2e^4}{\hbar^3}\sum_{\mathbf{k}}\sum_{\substack{n\in\text{VB}\\ m\in\text{CB}\\ l\neq n}} \text{Im}\left[\text{tr}_n\left(J_z^{-1}P_n(\mathbf{A}^*_{\omega_{mn}}\cdot\mathbf{v})P_m(\mathbf{A}_{\omega_{mn}}\cdot\mathbf{v}) \times \frac{P_n\left(j_y^{L_z} - i\hbar v_x/2 + j_y^{S_z}\right)P_l v_x}{\omega_{ln}^2}\right)\right], \quad (3)$$

where $\mathrm{j}^O \equiv \{O, \mathrm{v}\}/2$ represents the orbital (spin) current operator for $O = L_z$ ($O = S_z$), naturally recovering the standard Kubo formula for the orbital (spin) Hall conductivity [7,16]. Here, the additional term $-i\hbar v_x/2$ arises from the commutation relation between $L_z$ and the velocity operators $v_x$ and $v_y$, which we thus include as part of the orbital contribution. For details on the derivation of the formulae, see Supplemental Material [49].

Figure 4(b) shows the calculated Hall conductivity as a function of $\hbar\omega_{\mathrm{pump}}$. The prominent peak observed slightly above the bandgap is well reproduced by the theory, supporting our assignment of the signal to the photoexcited holes. Due to the selective injection of OAM polarization and the weak SOC in GaN, the orbital contribution is much larger in amplitude than the spin contribution. Crucially, the orbital and spin contributions possess opposite signs, with the orbital component (and thereby the total Hall conductivity) yielding a sign consistent with the experiment. This result strongly indicates that the observed Hall conductivity is a manifestation of the IOHE, namely, the conversion of OAM polarization into a transverse charge current. At the injected carrier density of $1.2 \times 10^{17}$ cm$^{-3}$ and the pump photon energy of 3.42 eV in our experiment, the calculated light-induced Hall conductivity is expected to reach ~ 1.0 S cm$^{-1}$ at $t_{\mathrm{pump}} =$ 0.3 ps. However, the experimental result in Fig. 4(a) is notably smaller, a possible reason of which is discussed later.

Figures 4(c)–4(e) plot the obtained values of $\tau_{yx}$, $C_{\mathrm{sj}}$, and $\omega_{\mathrm{skew}}$ in the fitting, which show no sharp dependence on $\hbar\omega_{\mathrm{pump}}$. The scatter away from the band gap (shaded regions) can be attributed to a degraded signal-to-noise ratio resulting from the small Hall signal. The transverse impurity scattering rate for OAM transport, $1/(2\pi\tau_{yx})$ ~ 0.3 THz, is found to be lower than the longitudinal momentum relaxation rate, $1/(2\pi\tau_{xx})$ ~ 1.1 THz. This is reasonable because while all photocarriers contribute to the longitudinal momentum relaxation ($\tau_{xx}^{-1}$), the IOHE signal ($\tau_{yx}^{-1}$) arises largely from the OAM-polarized holes which experience fewer scattering events near the $\Gamma$ point due to the limited channels. Intriguingly, $C_{\mathrm{sj}}$, the ratio of the side-jump to the intrinsic mechanisms, is close to $-2$, a well-established value for the anomalous and spin Hall effects in a simple parabolic band when incorporating the anomalous distribution [69,76]. Compared to $\Delta\sigma_{yx}^{\mathrm{int}}$ in Fig. 4(a), the extracted extrinsic parameters, in particular $\omega_{\mathrm{skew}}$, exhibit larger fluctuations, which can be accounted for by a degraded signal-to-noise ratio at the low-frequency part of our spectral window where the extrinsic contributions emerge.

**D. Dynamics of OAM transport**

To extract the ultrafast dynamics of the IOHE, we monitor $\Delta E_y(t_{\text{pump}})$ at the peak of the THz waveform ($t_{\text{probe}} =$ 1.9 ps). A typical temporal profile is shown in Fig. 5(a). Fitting the data with an exponential decay convolved with a Gaussian pump-pulse profile yields a time constant of $\tau_{\text{OAM}} \sim$ 0.5 ps. Because the $\Delta E_y$ signal tracks the evolution of the OAM polarization, this result provides an unprecedentedly direct time-domain measurement of the OAM transport governing the conversion to charge current within the single GaN layer. The inset summarizes the extracted values of $\tau_{\text{OAM}}$ at different pump photon energies. Strikingly, the observed decay occurs rapidly on a timescale comparable to the previously reported hot-hole relaxation time (0.6 ps at room temperature) [77].

Some earlier experiments on OAM transport reported that OAM lifetimes and diffusion lengths can be significantly longer than those of spin [17,20,22-25], which was theoretically ascribed to the near-degeneracy of states with different orbital characters [78]. However, recent theoretical developments [79-81] suggest that OAM—a characteristic closely tied to momentum-space texture—is actually less robust, sparking an ongoing debate. Corroborating the latter view, several recent experiments have indicated that the OAM diffusion length in metals is much shorter, even down to a few atomic layers [18,19,33]. Our present result in Fig. 5 provides a direct time-domain observation of OAM-polarized carrier transport in GaN, unambiguously demonstrating that OAM polarization is rapidly lost during the energy relaxation of carriers. Assuming a hole diffusion coefficient of $D_h \sim$ 0.85 $\text{cm}^2\ \text{s}^{-1}$ [82], the OAM relaxation length $\sqrt{D_h \tau_{\text{OAM}}}$ is approximately 6 nm.

Furthermore, the discrepancy between the observed intrinsic Hall conductivity in Fig. 4(a) and the prediction of the microscopic theory in Fig. 4(b)—with the observation being approximately 40 times smaller—implies the existence of an additional path for OAM relaxation which is even faster than the time resolution of the present experiment ($\sim$ 140 fs, limited by the UV pump pulse duration). Recent first-principles studies on unstrained GaN have suggested that holes are scattered primarily by LA phonons within 4 fs at room temperature [83,84]. If one assumes the OAM to be lost by this process, the true $\Delta E_y$ signal in the time domain would be suppressed by a factor of 35 under the finite time resolution. This may quantitatively account for the suppression of the observed signal relative to the microscopic calculation, yielding a much shorter OAM relaxation length of 0.6 nm. In contrast, the measured longer time constant ($\tau_{\mathrm{OAM}} \sim$ 0.5 ps) close to that of hole energy relaxation is most likely driven by LO phonon scattering [77].

As visualized in Figs. 5(b)-5(e), the light-induced OAM polarization of holes is highly confined to the $k_z$-axis and the $\Gamma$ point, where OAM is a good quantum number (albeit an approximate one due to weak but finite SOC). This momentum-space profile suggests a plausible mechanism for the two distinct timescales in the OAM relaxation as follows: The initial 4-fs decay represents quasi-elastic momentum redistribution of holes by low-energy LA phonons [83,84]. This process scatters the holes in the large-$k$ region outward from the $k_z$-axis, resulting in the loss of their OAM polarization; as the interband coherence is rapidly destroyed, they settle into local band eigenstates that no longer retain their OAM. However, such scattering does not significantly affect the holes near the band edge at the zone center. The LA phonons involved possess energies much lower than those of the holes [83] due to their velocity mismatch, and thus lack the momentum to scatter holes far from the $\Gamma$ point. Thus, the holes remaining around the zone center would keep their OAM until the subsequent scattering with LO phonons on a sub-picosecond time scale [77]. This inelastic scattering can move those holes away from the $\Gamma$ point or even induce inter-valence-band transitions, erasing their OAM that survived the initial LA phonon scattering. This scenario could also account for the insensitivity of $\tau_{\mathrm{OAM}}$ to $\hbar\omega_{\mathrm{pump}}$ [inset of Fig. 5(a)], since this slower relaxation is dominated by holes near the zone center regardless of the pump photon energy. It is worth noting that such momentum-resolved discussion is a unique advantage of our semiconductor-based optical approach, where the physical character of the involved carriers is unambiguously defined.

We point out that such a short OAM diffusion length itself is by no means a drawback for orbitronic applications. Achieving maximum OAM-to-current conversion efficiency within extremely thin films is highly advantageous for overall device miniaturization, effectively preventing inter-element crosstalk by strictly confining the interaction to the localized junction interfaces in real devices. Therefore, strategic selection of materials for long-range OAM transport and for local conversion based on their properties is crucial for realizing high performance. To enable this, time-domain quantitative evaluation of the OAM relaxation in each material is imperative. Our method is in principle applicable to various materials as long as interband optical transitions are induced by CP light pulses, opening a new avenue for ultrafast, non-contact characterization of spin and OAM transport.

## IV. Conclusion

In summary, we have successfully observed the ultrafast inverse orbital Hall effect in GaN using a CP pump and polarization-resolved THz spectroscopy. By circumventing the complexities associated with surfaces and interfaces—where the behavior of OAM remains poorly understood and actively debated—our non-contact THz transmission approach serves as a powerful platform for probing the pure bulk dynamics of orbital transport. Our experimental results and microscopic calculations clearly demonstrate that the observed transverse response originates primarily from the light-induced OAM polarization, providing a direct measurement of the orbital Hall conductivity. Crucially, by resolving the complex THz Hall conductivity spectrum, we quantitatively revealed a substantial contribution from extrinsic processes arising from asymmetric impurity scattering in the dc limit. These findings offer vital clues for constructing comprehensive microscopic models that encompass both intrinsic and extrinsic mechanisms. Furthermore, we found the OAM lifetime in GaN to be on a sub-picosecond timescale—or possibly even much shorter—a pioneering observation demonstrating the highly transient nature of OAM in a weak SOC regime without heterostructures. Ultimately, elucidating and controlling such extremely rapid relaxation processes is the crucial next step toward harnessing macroscopic OAM in semiconductors, advancing the development of sustainable, next-generation spintronic and orbitronic technologies.

## Acknowledgement

The authors acknowledge Yuki Kamitani for his help in sample fabrication. This work was supported by JST FOREST (Grant No. JPMJFR2240), JSPS KAKENHI (Grant Nos. JP24K00550 and JP24K16988), Toray Science Foundation (Grant No. 25-6609), JST CREST (Grant No. JPMJCR20R4), and MEXT Q-LEAP (Grant No. JPMXS0118068681). Infrared and UV transmission measurements were performed using the facilities of Materials Design and Characterization Laboratory in The Institute for Solid State Physics, The University of Tokyo.

R.M. conceived the project. M.F. and Y.K. prepared the sample. K.A., T.F., A.M.S., and Y.M. developed the pump-probe spectroscopy system with the help of J.Y. and R.M. K.A. performed the experiment, analyzed the data, and interpreted the results with the help of T.F., A.M.S., Y.M., S.M., and R.M. K.A. performed the theoretical calculations with the help of T.F. and Y.M. All the authors discussed the results. K.A. and R.M. wrote the manuscript with substantial feedback from S.M. and Y.M. and coauthors.

## Figures

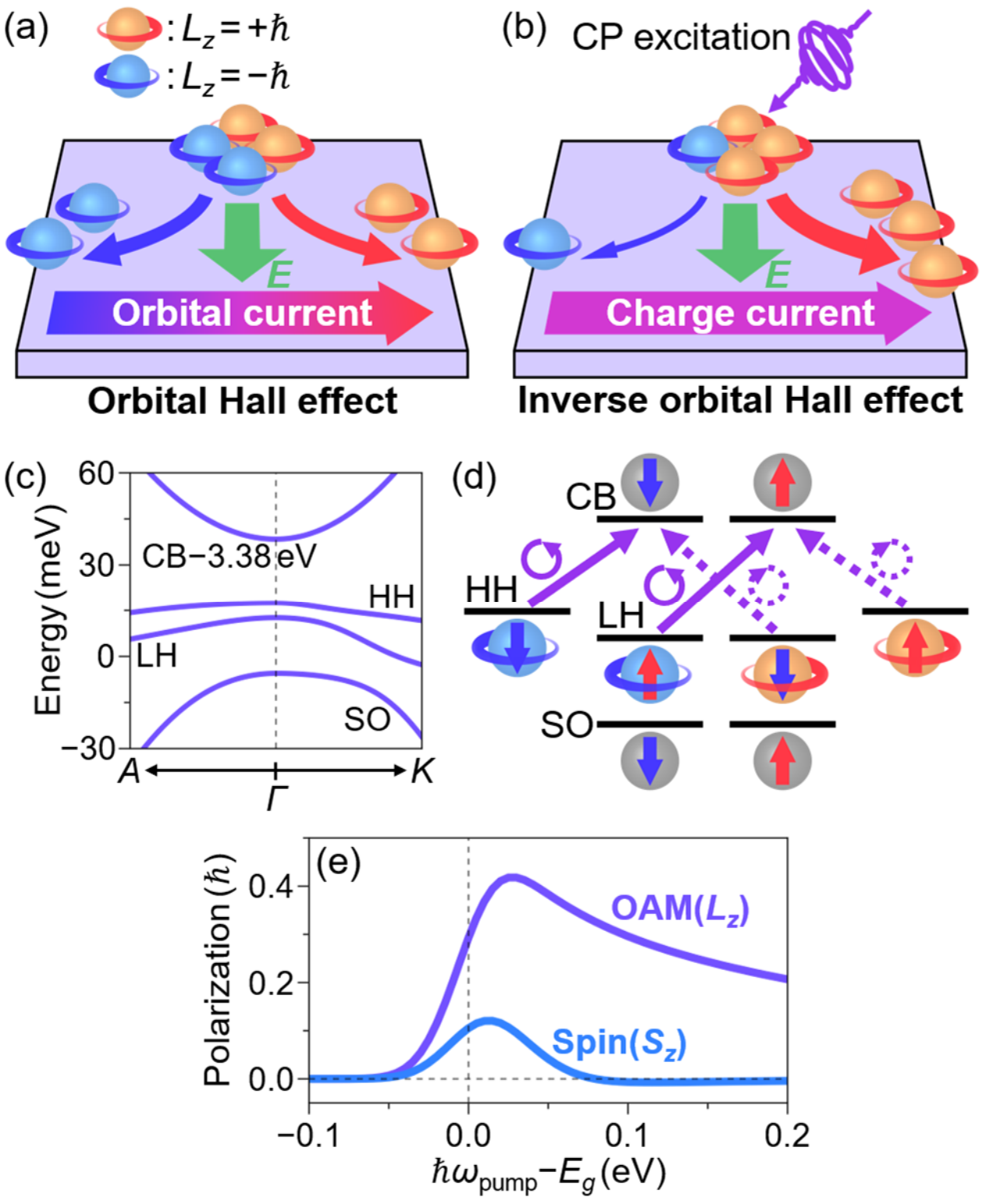

FIG. 1. (a,b) Schematics of the orbital Hall effect (a) and the light-induced inverse orbital Hall effect (b). (c) Band structure of GaN near the band edge, consisting of the conduction band (CB) and the heavy-hole (HH), light-hole (LH), and split-off (SO) bands. (d) Approximate selection rules for the interband transitions at the zone center in GaN. Violet solid (dotted) arrows indicate allowed transitions under photoexcitation by left(right)-handed CP light. (e) $\hbar\omega_{\mathrm{pump}}$ dependence of calculated light-induced OAM and spin polarizations, defined as the expectation value of each operator for excited holes. The results are normalized by the pump photon density.

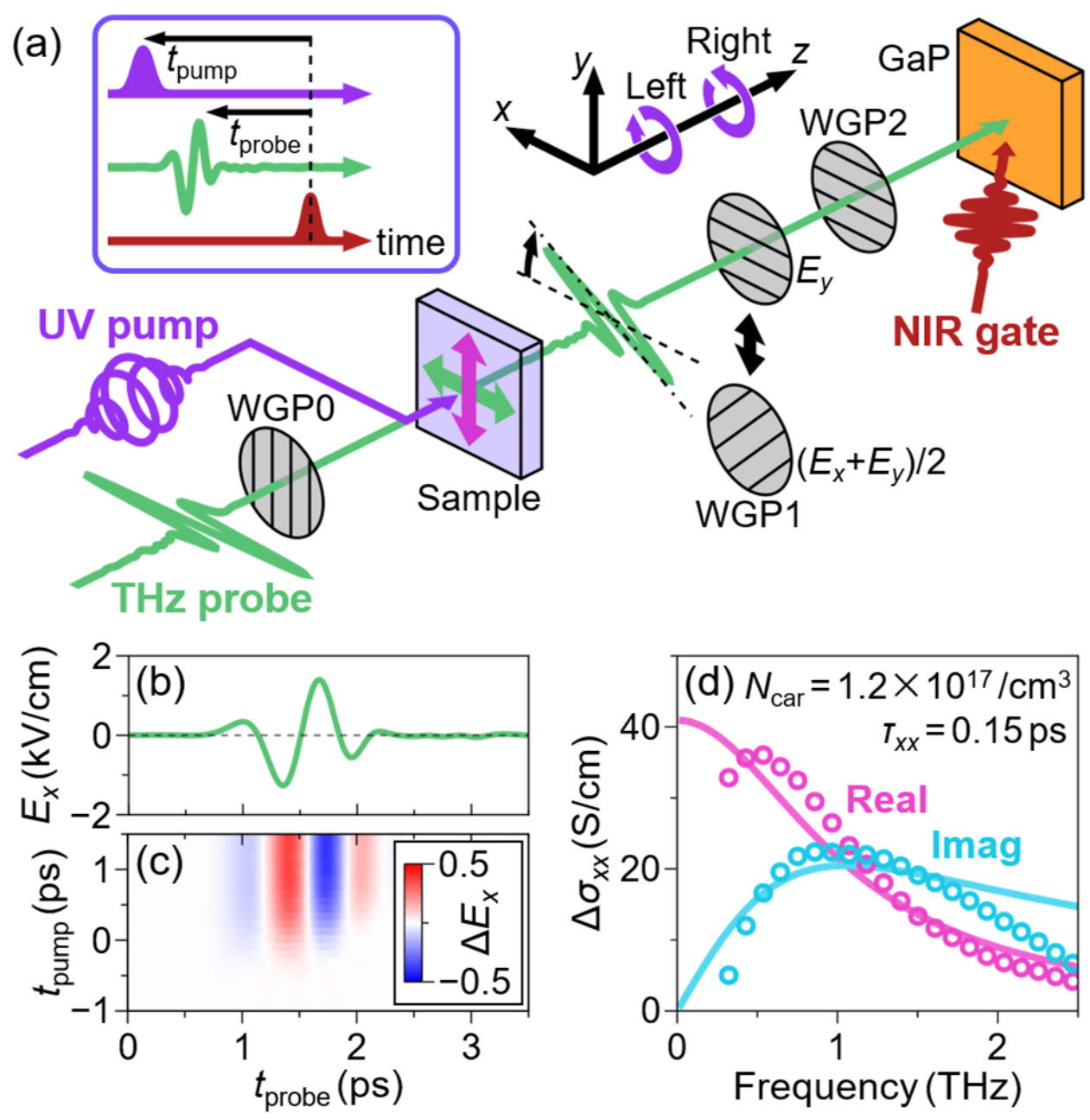

FIG. 2. (a) Schematic of the THz polarimetry measurement using UV CP pump (violet), THz probe (green) and NIR gate (red) pulses. The inset shows the pulse sequence. (b) Transmitted THz waveform $E_x(t_{\mathrm{probe}})$ in the absence of the pump pulse. (c) 2D plot of the light-induced change in the THz field $\Delta E_x(t_{\mathrm{pump}}, t_{\mathrm{probe}})$. (d) Light-induced longitudinal conductivity spectrum $\Delta\sigma_{xx}(\omega)$ at $t_{\mathrm{pump}} =$ 4 ps, alongside the Drude model fit (solid curves). The data in (c,d) were taken at $\hbar\omega_{\mathrm{pump}} =$ 3.42 eV and a pump fluence of 100 μJ cm$^{-2}$.

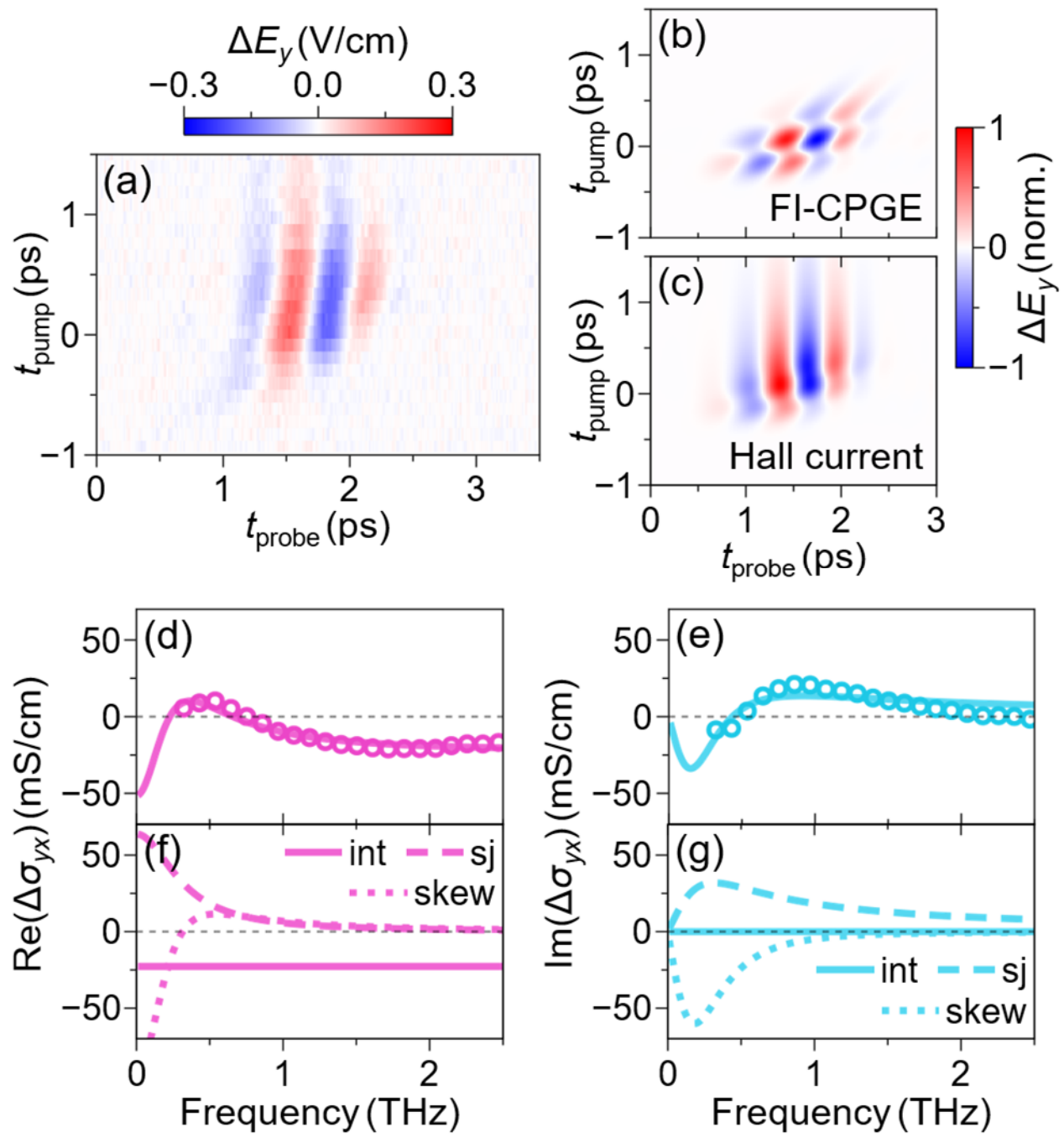

FIG. 3. (a) 2D plot of the helicity-dependent THz field $\Delta E_y(t_{\mathrm{pump}}, t_{\mathrm{probe}})$. (b,c) Calculated 2D profiles of $\Delta E_y(t_{\mathrm{pump}}, t_{\mathrm{probe}})$ originating from the FI-CPGE (b) and the anomalous Hall current of photocarriers (c) (see Supplemental Material [49]). (d,e) Real (d) and imaginary (e) parts of the light-induced Hall conductivity spectrum $\Delta\sigma_{yx}(\omega)$ at $t_{\mathrm{pump}} =$ 0.3 ps, alongside a phenomenological model fit (solid curves). (f,g) Intrinsic (int), side-jump (sj) and skew scattering (skew) contributions to the phenomenological fit in (d,e). The data in (a,d,e) were taken at $\hbar\omega_{\mathrm{pump}} =$ 3.42 eV and a pump fluence of 100 μJ cm$^{-2}$.

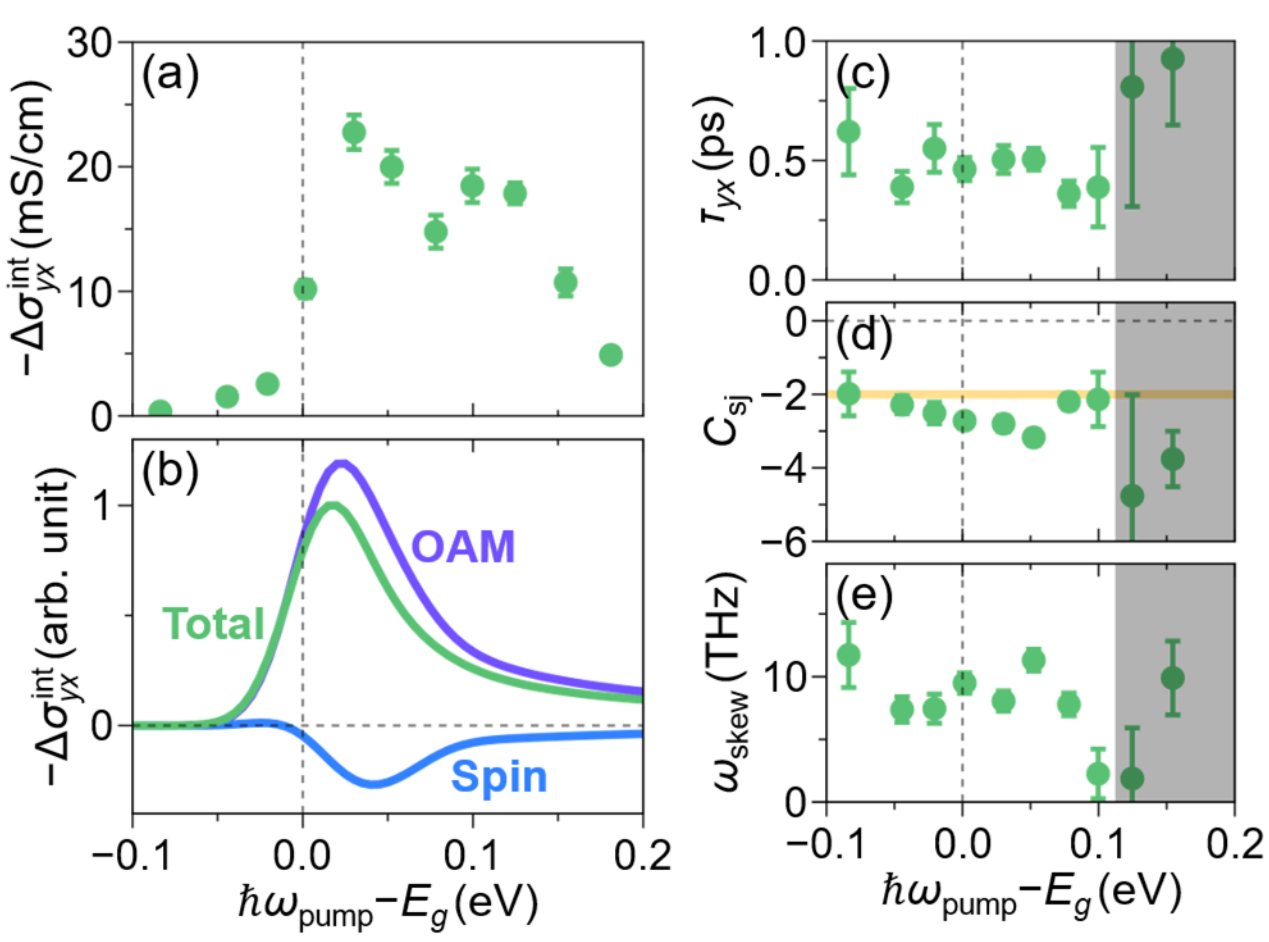

FIG. 4. (a) $\hbar\omega_{\rm pump}$ dependence of $\Delta\sigma_{yx}^{\rm int}$ extracted from light-induced Hall conductivity spectra at $t_{\rm pump} =$ 0.3 ps and a pump fluence of 100 μJ cm$^{-2}$. (b) Calculated $\hbar\omega_{\rm pump}$ dependence of light-induced anomalous Hall conductivity (labeled "Total"), alongside the orbital and spin contributions. (c-e) $\hbar\omega_{\rm pump}$ dependence of the fitted parameters for $\tau_{yx}$ (c), $C_{\rm sj}$ (d), and $\omega_{\rm skew}$ (e) (see text). The orange line in (d) indicates $C_{\rm sj} = -2$. The shaded regions indicate a high-noise photon energy range due to the small Hall response. The error bars represent the standard error derived from the phenomenological fit.

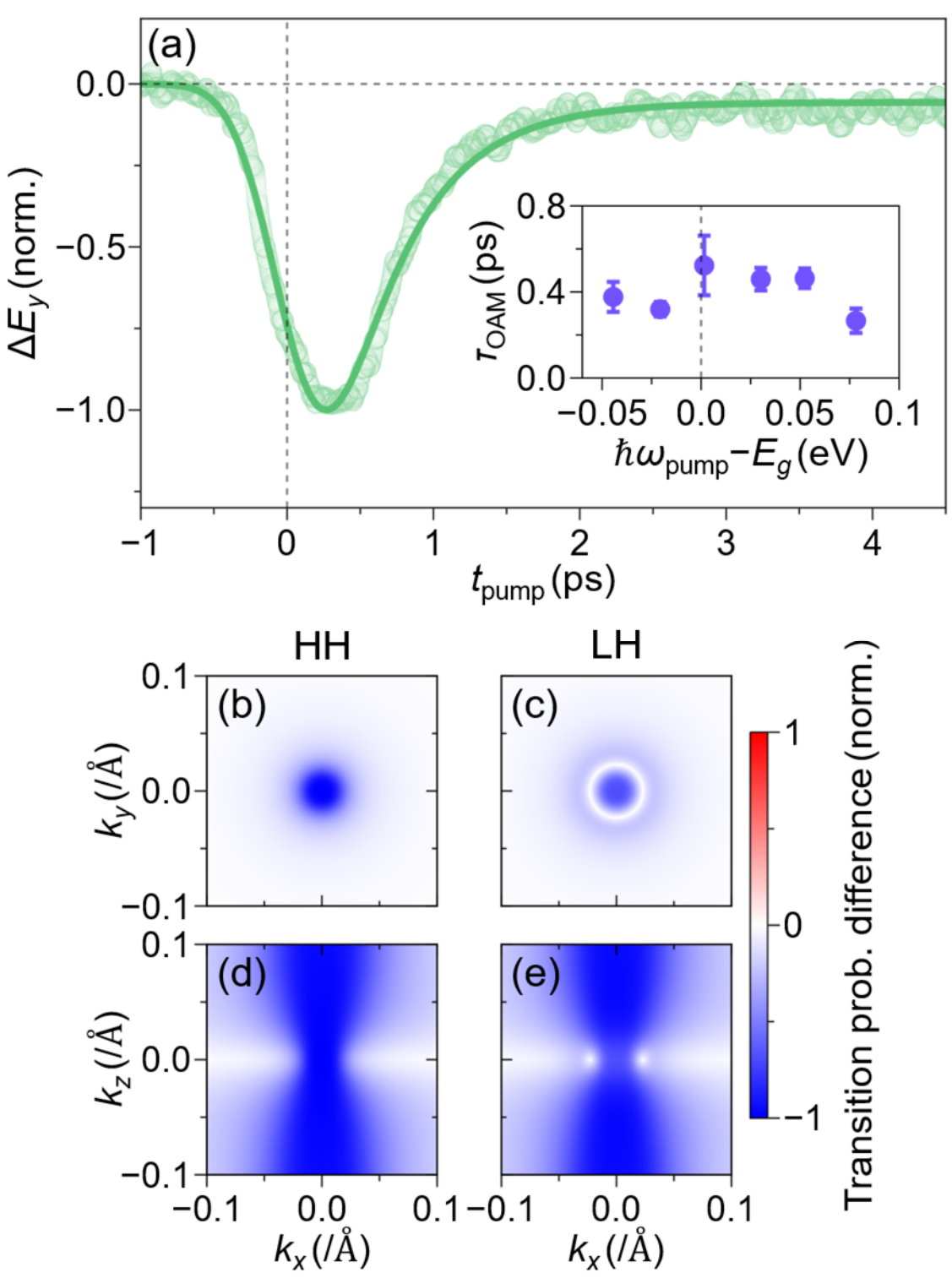

FIG. 5. (a) THz field $\Delta E_y$ as a function of the UV pump delay $t_{\mathrm{pump}}$ at $t_{\mathrm{probe}} =$ 1.9 ps and $\hbar\omega_{\mathrm{pump}} =$ 3.42 eV, alongside an exponentially modified Gaussian fit (solid curve). The inset shows the values of $\tau_{\mathrm{OAM}}$ extracted at various $\hbar\omega_{\mathrm{pump}}$. The data were taken at a pump fluence of 100 μJ cm$^{-2}$. (b-e) Transition probability difference between Kramers-degenerate states in the HH (b,d) and LH (c,e) bands under photoexcitation by spectrally uniform left-handed CP light, mapped in the $xy$-plane (b,c) and $xz$-plane (d,e). The overall negative sign indicates the dominance of transitions from states with negative $L_z^{\Gamma}$, resulting in the injection of holes with positive OAM.